\documentstyle[12pt]{article}

\newcommand{\bd}{\begin{document}}
\newcommand{\ed}{\end{document}}
\newcommand{\bc}{\begin{center}}
\newcommand{\ec}{\end{center}}
\newcommand{\be}{\begin{eqnarray}}
\newcommand{\ee}{\end{eqnarray}}
\renewcommand{\thefootnote}{\alph{footnote}}
\newcommand{\se}{\section}
\newcommand{\sse}{\subsection}
\newcommand{\bi}{\bibitem}
\def\figcap{\section*{Figure Captions\markboth
     {FIGURECAPTIONS}{FIGURECAPTIONS}}\list
     {Figure \arabic{enumi}:\hfill}{\settowidth\labelwidth{Figure 999:}
     \leftmargin\labelwidth
     \advance\leftmargin\labelsep\usecounter{enumi}}}
\let\endfigcap\endlist \relax

\begin{document}

\begin{titlepage}

 \vskip 0.5in
 \null
\begin{center}
 \vspace{.15in}
{\Large {\bf CP Violation in Hyperon Decays from SUSY \\with
Hermitian Yukawa and $A$ Matrices
}}\\
\vspace{1.0cm}  \par
 \vskip 2.1em

{\Large\bf Chuan-Hung Chen}

\vspace{0.7cm}

   {\Large \sl Institute of Physics, Academia Sinica, Taipei, }\\
   { \Large \sl Taiwan 115,
Republic of China \\}

 \par \vskip 5.3em

\date{\today}
 {\Large\bf Abstract}

\end{center}
We show that a large CP asymmetry in hyperon decays can be naturally
realized in the framework of SUSY models. The possibility is
implemented by the hermiticities of Yukawa and  $A$ matrices. And
also, the observed values of $\epsilon$ and $\epsilon'$ are
explained.
\end{titlepage}

The origin of CP violation (CPV) is one of mysterious phenomena in particle
physics since its discovery in the neutral kaon decays in 1964 \cite{CCFT}.
Recently, another CP asymmetry, $\sin 2\beta$, in the decay of $B\rightarrow
J/\Psi K_{s}$ is observed by BARBAR \cite{betaBABAR} and BELLE \cite
{betaBELLE}. Nevertheless, our understanding of CPV is still exiguous.\ In
the standard model (SM), the unique source of CPV is from the
Cabibbo-Kobayashi-Maskawa (CKM) matrix \cite{CKM} induced from the
three-generation quark mixings and described by the three angles $\alpha $, $%
\beta $ and $\gamma $ or $\phi _{2}$, $\phi _{1}$ and $\phi _{3}.$ Even
though the SM prediction on the indirect CP violating parameter $\epsilon $
in the kaon system can be fitted well with current experimental data, due to
the large uncertainties from hadronic matrix elements, so far it is unclear
whether the result in the SM is consistent with the observed value of the
direct CP violating parameter $\epsilon ^{\prime }$ measured by KTeV \cite
{KTeV} and NA31 \cite{NA31}. Furthermore, the problem of baryogenesis is
insolvable in the SM. In addition, the predicted CP asymmetry of $O(10^{-5})$
in hyperon decays of $\Xi\rightarrow \Lambda \pi \rightarrow p\pi \pi $ is
about one order of the magnitude smaller than that of $O(10^{-4})$ proposed
by the experiment E871 at Fermilab \cite{White}. Hence, it is inevitable to
look for new physics which gives observable CPV effects.

Supersymmetric theories not only supply an elegant mechanism for the
breaking of the electroweak symmetry and a solution to the hierarchy
problem, but provide many new weak CP violating phases. These CP phases
usually arise from the trilinear and bilinear supersymmetry (SUSY) soft
breaking $A$ and $B$ terms, the $\mu $ parameter for the scalar mixing and
gaugino masses, respectively. Unfortunately, it has been shown that with the
universal assumption on soft breaking parameters, these phases are severely
bounded by electric dipole moments (EDMs) \cite{Garistosusy} so that the
contributions to $\epsilon $ and $\epsilon ^{\prime }$ are far below the
experimental values. In the literature, some strategies to escape the
constraints of EDMs have been suggested. They are mainly (a) by setting the
squark masses of the first two generations to be as heavy as few TeV \cite
{BKMW} but allowing the third one being light; (b) by including all possible
contributions to EDMs such that somewhat cancellations occur in some allowed
parameter space \cite{IN,BGK}; and (c) with the non-universal soft $A$ terms
instead of universal ones. In particular, those models with non-universal
parameters have been demonstrated that they can be realized in some
string-inspired models \cite{String1,String2,String3,String4}.

Among the models with non-degenerate soft trilinear terms, for satisfying
the bounds of EDMs, the phases in the diagonal elements of the $A$ matrix
should be set to be small in any basis artificially although the remaining
large phases and the light sfermions are still allowed to explain the
observed values of $\epsilon $ and $\epsilon ^{\prime }$. To overcome this
problem, it is proposed in Ref. \cite{ABKL} to use hermitian Yukawa and $A$
matrices. The construction of a hermitian Yukawa matrix can be implemented
based on some symmetries such as the global (gauged) horizontal $SU(3)_{H}$
symmetry \cite{MY} and left-right symmetry \cite{BDM}. Although the
hermiticity will be broken by renormalization group (RG) effects, it is
shown \cite{ABKL} that their contributions to EDMs are two orders of the
magnitude below the present experimental limit. Moreover, due to the
hermitian property, a special relation is obtained as
\begin{equation}
\left( \delta _{12}^{d}\right) _{LR}\simeq \left( \delta _{12}^{d}\right)
_{RL}  \label{mi}
\end{equation}
where $(\delta _{12}^{d})_{LR}\equiv (V^{d\dagger }A^{d}v_{d}V^{d})_{12}/m_{\tilde{q%
}}^{2}$, $A^{d\dagger }\simeq A^{d}$, $v_{d}$ is the vacuum expectation
value (VEV) of the Higgs filed $\Phi _{d}$ for supplying the masses of
down-type quarks , $V^{d}$ is the mixing matrix for diagonalizing the mass
matrix of down-type quarks and $m_{\tilde{q}}$ is the average mass of squark
in super-CKM basis. We note that the mixing matrix for mass eigenstates of
left-handed down-type quarks is the same as that for the right-handed one.
In this paper, we will show the implication of Eq. (\ref{mi}) on the CP
asymmetry of hyperon decays.

The interactions describing $|\Delta S|=1$ nonleptonic decays of $\Xi $ and $%
\Lambda $ are the same as those for $K\rightarrow \pi \pi $ processes.
Therefore, those CP violating effects contributing to hyperon decays will
also contribute to $\epsilon ^{\prime }$. As a consequence, the CP
observable for hyperon decays is limited to be $O(10^{-5})$ \cite{He,CC}
level by the bound of $\epsilon ^{\prime }$. One way to avoid the constraint
is that the couplings contributing to parity conserving parts of hyperon
decays are enhanced but suppressed for parity violating ones.

To understand the CPV in hyperon decays, we start by writing the decay
amplitude as \cite{MRR}
\begin{equation}
Amp(B_{i}\to B_{f}\pi )=S+P\vec{\sigma}\cdot \hat{q}  \label{amp}
\end{equation}
where $B_{i,f}$ are the initial and final baryons, $S$ and $P$ denote the
parity violating and conserving amplitudes, respectively, and $\hat{q}$ is
the momentum direction of outgoing baryon $B_{f}$. We note that Eq. (\ref
{amp}) has to be multiplied by a factor of $G_{F}m_{\pi }^{2}$, with $m_{\pi
}$ being the pion mass, for getting correct decay rate. For simplicity, the
amplitudes $S$ and $P$ can be parametrized as
\begin{eqnarray}
S &=&\sum_{i}S_{i}e^{i(\delta _{i}^{S}+\theta _{i}^{S})},  \nonumber \\
P &=&\sum_{i}P_{i}e^{i(\delta _{i}^{P}+\theta _{i}^{P})}  \label{eqn:sp}
\end{eqnarray}
where we have separated the strong phases $\delta _{i}$ generated by final
state interactions and the weak CP violating phases $\theta _{i}$ from decay
amplitudes such that $S_{i}$ and $P_{i}$ amplitudes are real, with $i$
representing all possible final isospin states. The decay distribution of
proton for the chain decays $\Xi \rightarrow \Lambda \pi \rightarrow p\pi
\pi $ with unpolarized $\Xi $ is then given by \cite{MRR}
\begin{equation}
4\pi \frac{dP}{d\Omega }=1+\alpha _{\Lambda }\alpha _{\Xi }\hat{p}_{\Lambda
}\cdot \hat{p}  \label{pd}
\end{equation}
and
\[
\alpha _{H}\equiv 2Re(S_{H}^{*}P_{H})/(|S_{H}|^{2}+|P_{H}|^{2})
\]
where $\alpha _{\Xi (\Lambda )}$ is the polarization of $\Lambda (p)$ for $%
\Xi \rightarrow \Lambda \pi $ $(\Lambda \rightarrow p\pi )$. According to
Eq. (\ref{pd}), the direct CP violating observable ${\cal A}$ can be defined
as:
\begin{eqnarray}
{\cal A} &=&\frac{\alpha _{\Xi }\alpha _{\Lambda }+\bar{\alpha}_{\Xi }\bar{%
\alpha}_{\Lambda }}{\alpha _{\Xi }\alpha _{\Lambda }-\bar{\alpha}_{\Xi }\bar{%
\alpha}_{\Lambda }},  \nonumber \\
&\approx &{\cal A}_{\Xi }+{\cal A}_{\Lambda }
\end{eqnarray}
with
\[
{\cal A}_{H}=\frac{\alpha _{H}+\bar{\alpha}_{H}}{\alpha _{H}-\bar{\alpha}_{H}%
}\ \ (H=\Xi \,,\ \Lambda )
\]
where $\bar{\alpha}_{H}$ is the corresponding quantity for the antihyperon $H
$. Although the $|\Delta S|=1$ hyperon decays include two isospin channels $%
\Delta I=1/2$ and $\Delta I=3/2$,
the contribution of $\Delta I=3/2$ amplitude can be neglected so that the
asymmetries for $\Xi \to \Lambda \pi $ and $\Lambda \to P\pi $, from Eq. (%
\ref{eqn:sp}), can be obtained as \cite{DHP,He,CC}
\begin{eqnarray}
{\cal A}_{\Xi } &\approx &-tan(\delta _{2}^{P}-\delta _{2}^{S})sin(\theta
_{2}^{P}-\theta _{2}^{S}),  \nonumber \\
{\cal A}_{\Lambda } &\approx &-tan(\delta _{1}^{P}-\delta
_{1}^{S})sin(\theta _{1}^{P}-\theta _{1}^{S}).
\label{eqn:asylambda}
\end{eqnarray}
It is known that the strong phases for $\Lambda \rightarrow p\pi $
decay are $\delta _{1}^{S}=6.0^{o}$ and $\delta _{1}^{P}=-1.1^{o}$
\cite{Roper}. However, for the $\Xi $ decay, we take $\delta
_{2}^{S}=0.2^{o}$ and $\delta _{2}^{P}=-1.7^{o}$ calculated by
using the chiral perturbation theory \cite {LSW}. The result of
$\delta _{2}^{S}$ recently is confirmed in the framework of a
relativistic chiral unitary approach \cite{MO}. Due to the small values of $%
\delta _{2}^{S}$ and $\delta _{2}^{P}$, consequently, the CP asymmetry of $%
\Xi $ is smaller than that of $\Lambda $ by one order of the magnitude.
Hence, in our following analysis, we only concentrate on ${\cal A}_{\Lambda }
$. 
To estimate the weak CP violating phases $\theta _{1}^{P}$ and $\theta
_{1}^{S}$, we adopt the following approximation \cite{He}
\begin{equation}
\theta _{1}^{l}\approx {\frac{Im[{\cal M}(\Lambda \to p\pi )|_{1}^{l}]}{Re[%
{\cal M}(\Lambda \to p\pi )|_{1}^{l}]}}  \label{eqn:theta1}
\end{equation}
by assuming that the CP violating contributions are much less than the CP
conserving ones. Here, ${\cal M}(\Lambda \to p\pi )$ express the transition
matrix elements of relevant effective operators, and their real parts can be
obtained from the experimental measurements, with the parity violating and
conserving amplitudes of $l=0$ and $l=1$, respectively. In sum, CP violating
phases for $\Lambda \rightarrow p\pi $ decay can be written as
\begin{eqnarray}
\theta _{1}^{S} &\approx &{\frac{Im[{\cal M}(\Lambda \to p\pi )|_{1}^{S}]}{%
1.47G_{F}m_{\pi }^{2}}},  \label{eqn:thetas} \\
\theta _{1}^{P} &\approx &-{\frac{Im[{\cal M}(\Lambda \to p\pi )|_{1}^{P}]}{%
9.98G_{F}m_{\pi }^{2}}}  \label{eqn:thetap}
\end{eqnarray}
where we have used $S_{1}=1.47$ and $P_{1}=0.6.$

As stated early, the interactions for $\Lambda \to p\pi $ are the same as
those for $K\to \pi \pi $. According to the analysis of Refs. \cite{GGMS,MM}%
, in SUSY models the main effects for $\epsilon ^{\prime }$ are from the
gluino penguin contributions. The associated effective Lagrangian is given
by \newline
\begin{equation}
{\cal L}_{eff}=C_{8}(\mu )O_{8}+\tilde{C}_{8}(\mu )\tilde{O}_{8}
\label{heff}
\end{equation}
where the effective Wilson coefficient $C_{8}(\mu )$ and the operator $O_{8}$
are expressed by
\begin{eqnarray}
O_{8} &=&\frac{g_{s}m_{s}}{8\pi ^{2}}\bar{d}\,i\sigma _{\mu \nu
}t^{a}P_{R}\,s\,G_{a}^{\mu \nu }\,,  \nonumber \\
C_{8}(\mu ) &=&{\frac{\alpha _{s}\pi }{m_{\tilde{q}}^{2}}}\frac{m_{\tilde{g}}%
}{m_{s}}\left( \delta _{12}^{d}\right) _{LR}(-\frac{1}{3}M_{1}+3M_{2})\,,
\label{c8}
\end{eqnarray}
respectively, with
\begin{eqnarray*}
M_{1}(x) &=&\frac{1+4x-5x^{2}+4x\ln (x)+2x^{2}\ln (x)}{2\left( 1-x\right)
^{4}}, \\
M_{2}(x) &=&x^{2}\frac{5-4x-x^{2}+2\ln (x)+4x\ln (x)}{2\left( 1-x\right) ^{4}%
}.
\end{eqnarray*}
$\tilde{O}_{8}$ and $\tilde{C}_{8}(\mu )$ in Eq. (\ref{heff}) can be
obtained from $O_{8}$ and $C_{8}(\mu )$ easily by changing the role of
chirality therein each other, $(\delta _{12}^{d})_{LR(RL)}$ denotes the
mixing effect between left (right)- and right (left)-handed squarks, $m_{%
\tilde{g}}$ is the gluino mass, $Tr(t^{a}t^{b})=\delta ^{ab}/2 $ and $x=m_{%
\tilde{g}}^{2}/m_{\tilde{q}}^{2}$. From Eq. (\ref{c8}), we see that this
interaction is no further suppression from the light quark mass.

In terms of Eq. (\ref{heff}), we know that $\epsilon ^{\prime }$ will be
related to $Im\left[ (\delta _{12}^{d})_{LR}-(\delta _{12}^{d})_{RL}\right] $
in which minus is from the different chirality. In general, it is not
necessary that $(\delta _{12}^{d})_{LR}$ is the same as $(\delta
_{12}^{d})_{RL}$. Therefore, it is often concluded that SUSY models can
agree with the measured value of $\epsilon ^{\prime }$. As for the hyperon
CPV, from Eqs. (\ref{eqn:thetas}) and (\ref{eqn:thetap}), we get $\theta
_{1}^{s}\propto Im\left[ (\delta _{12}^{d})_{LR}-(\delta
_{12}^{d})_{RL}\right] $ and $\theta _{1}^{p}\propto -Im\left[ (\delta
_{12}^{d})_{LR}+(\delta _{12}^{d})_{RL}\right] $. If we assume that $\theta
_{1}^{s}$ is the dominant one, due to the constraint of $\epsilon^{\prime}$,
the CPV of $O(10^{-5})$ in Eq. (\ref{eqn:asylambda}) can be obtained. But,
if we set $\theta _{1}^{p}$ to be the dominant one, $\epsilon ^{\prime }$
from the same mechanism will be suppressed. It is interesting to ask whether
$\epsilon ^{\prime }/\epsilon\sim 2\times 10^{-3}$ and ${\cal A}_{\Lambda
}\sim 10^{-4}$ can both be reached in the framework of SUSY.

From the above analysis, we know that it is hopeless if the mechanism for $%
\epsilon ^{\prime }$ and for ${\cal {A}}_{\Lambda }$ is the same. However,
the possibility can be realized if Yukawa and non-universal $A$ matrices are
hermitian. As mentioned before, such a kind of SUSY model implies $(\delta
_{12}^{d})_{LR}\simeq $ $(\delta _{12}^{d})_{RL}$. That is, $\epsilon
^{\prime }$ is suppressed in this gluino penguin contribution but $\theta
_{1}^{p}$ is enhanced. To emphasize that both large $\epsilon ^{\prime }$
and ${\cal A}_{\Lambda }$ can be obtained in the SUSY model, for simplicity,
we adopt the CP violating phase arises from the Yukawa matrix and other SUSY
parameters are real \cite{ABKL}. Although this SUSY model does not introduce
the new weak CP phases, it still provides an abundant particle spectrum and
flavor physics that can be tested in the current and future experiments. As
a result, W-boson and charginos will contribute to $\epsilon ^{\prime }$.
Similar to the SM, by combining these effects altogether, $\epsilon^{\prime}$
can be in agreement with the experimental value even though it is sensitive
to the uncertainties of hadronic matrix elements. It is shown that even
using only chargino box-diagram contributions \cite{KL}, with more generic
assumptions on SUSY models, the result is also possibly consistent with the
data.

As known, $\epsilon $ from gluino box diagrams can give a bound on $%
Im(\delta _{12}^{d})_{LR}$ through $Im(\delta _{12}^{d})_{LR}^{2}$ \cite
{GGMS}. In order to constrain $Im(\delta _{12}^{d})_{LR}$ directly, we
consider the contributions of long-distance effects to $\epsilon $, in which
the transition matrix element for $\left\langle \bar{K}^{0}\right|
H_{eff}\left| K^{0}\right\rangle $ comes from the $\pi ,$ $\eta $ and $\eta
^{\prime }$ poles. According to Ref. \cite{DH}, the result is shown as

\begin{equation}
\epsilon _{LD}\approx {\frac{\omega }{40\sqrt{2}(m_{K}^{2}-m_{\pi
}^{2})m_{K}\Delta m_{K}}}<K^{0}|{\cal L}_{even}|\pi ^{0}><\pi ^{0}|{\cal L}%
_{odd}|\bar{K}^{0}>  \label{epsilon}
\end{equation}
where $\Delta m_{K}$ is the mass difference of $K_{L}$ and
$K_{S}$, $\omega $ stands for the contributions from different
poles and its accessible range is $1<\left| \omega \right| <4$,
${\cal L}_{even(odd)}$ denotes the CP-even (odd) interaction and
the explicit expression of ${\cal L}_{odd}$ is
\[
{\cal L}_{odd}={Im}\left( f_{PC}\right) \bar{d}\,i\sigma _{\mu \nu
}t^{a}\,s\,G_{a}^{\mu \nu }
\]
with
\begin{eqnarray}
f_{PC} &=&\frac{g_{s}\alpha _{s}\eta _{g}}{16\pi m_{\tilde{g}}}\left[ \left(
\delta _{12}^{d}\right) _{LR}+\left( \delta _{12}^{d}\right) _{RL}\right] {x}%
(-\frac{1}{3}M_{1}(x)+3M_{2}(x))  \label{fpc} \\
\eta _{g} &=&\left( \frac{\alpha _{s}(\mu _{\Lambda })}{\alpha _{s}(m_{c})}%
\right) ^{-14/27}\left( \frac{\alpha _{s}(m_{c})}{\alpha _{s}(m_{b})}\right)
^{-14/25}\left( \frac{\alpha _{s}(m_{b})}{\alpha _{s}(m_{t})}\right)
^{-14/23}\left( \frac{\alpha _{s}(m_{t})}{\alpha _{s}(m_{\tilde{g}})}\right)
^{-14/21}  \nonumber
\end{eqnarray}
where $\eta _{g}$ is the QCD effects \cite{CHP}. For the CP-even part, we
can use the experimental value $<K^{0}|{\cal L}_{even}|\pi ^{0}>\approx
2.58\times 10^{-7}$ GeV$^{2}$. However, for the CP-odd part, according to
the MIT bag model \cite{DHH}, we have $<\pi ^{0}|{\cal L}_{odd}|\bar{K}%
^{0}>\approx Im(f_{PC})A_{K\pi }$ and $A_{K\pi }=0.4\ GeV^{3}$ for $\alpha
_{s}\approx 1$. Hence, the long-distance effects on $\epsilon $ is
\[
\epsilon _{LD}\approx {4.8\times 10}^{6}\omega {Im}f_{PC}.
\]
Requiring the value of $\epsilon _{LD}$ being less than
$2.28\times 10^{-3},$ the upper bound can be given as
$Imf_{PC}\leq \left( 4.7/\omega \right) \times 10^{-10}.$

Due to $(\delta _{12}^{d})_{LR}\simeq $ $(\delta _{12}^{d})_{RL}$ in our
case, the weak phase $\theta _{1}^{S}$ is negligible. By using Eq. (\ref
{eqn:thetap}) and the matrix element calculated by the MIT bag model \cite
{DHP,HMPV}, the CP violating phase $\theta _{1}^{P}$ can be given as
\[
\theta _{1}^{p}\approx -4.8\times 10^{6}{Im}(f_{PC})B_{p}
\]
where $B_{p}$ represents the uncertainty in estimating the matrix elements
of hyperon decays and the allowed range is $0.35<B_{p}<2.6$ \cite{HMPV}. In
terms of Eq. (\ref{eqn:asylambda}) and the bound of ${Im}f_{PC}$, we obtain
\[
|{\cal A}_{\Lambda }|\leq 2.93\times 10^{-4}\frac{B_{p}}{\left|
\omega \right| }.
\]
Although the result is sensitive to the theoretical uncertainty, by taking a
proper value, the CP asymmetry ${\cal A}_{\Lambda }$ can reach $O(10^{-4})$
easily.

In summary, it has an enormous progress in SUSY models since a nonzero value
of $\epsilon ^{\prime }$ is confirmed by the KTeV experiment. Although these
models can lead to the observed values of $\epsilon $ and $\epsilon ^{\prime
}$ well, with the same mechanism and without a further fine tuning, the
predicted CP asymmetry in hyperon decays is below the expected value
proposed by the E871 experiment. Hence, to obtain large values for both $%
\epsilon ^{\prime }$ and ${\cal A}_{\Lambda }$, the Feynman diagrams for
each of them should be different. We show that the observed value of $%
\epsilon ^{\prime }$ and ${\cal A}_{\Lambda }=O(10^{-4})$ can be reached in
the framework of SUSY models naturally if Yukawa and soft breaking A terms
are hermitian. In addition, once the CP asymmetry of $O(10^{-4})$ is
measured in hyperon decays, it also gives a strong evidence to support the
existence of SUSY.\newline

\noindent {\bf Acknowledgments}

I would like to thank D. Chang, X.G. He, C.Q. Geng and H.N. Li for
their useful discussions. This work was supported by the National
Science Council of the Republic of China under Contract
NSC-90-2112-M-001-069.

\end{document}